\documentstyle[epsfig,aps,prc]{revtex}

\begin{document}

\title{Supersymmetric Transformations in Coupled-Channel Systems
}
\author{H. Leeb$^1$, S. A. Sofianos$^2$, J.-M. Sparenberg$^3$, 
and D. Baye$^3$}
\address{$^1$Institut f\"ur Kernphysik, Technische Universit\"at Wien,
Wiedner Hauptstra\ss e 8-10/142, A-1040 Vienna, Austria}
\address{$^2$Department of Physics, University of South Africa,
P.O. Box 392, Pretoria 0003, South Africa}
\address{$^3$Physique Nucl\'eaire Th\'eorique et Physique Math\'ematique, 
C.P. 229, Universit\'e Libre de Bruxelles, B-1050 Brussels, Belgium} 
\date{\today}
\maketitle

\begin{abstract}
A  transformation of supersymmetric quantum mechanics for $N$ 
coupled channels is presented, which allows the introduction of up
to $N$ degenerate bound states without altering the remaining spectrum 
of the Hamiltonian. Phase equivalence of the Hamiltonian can be restored 
by two successive supersymmetric transformations at the same energy. The 
method is successfully applied to the $^3$S$_1$-$^3$D$_1$ coupled 
channels of the nucleon-nucleon system  and a set of Moscow--type 
potentials is thus generated.
\end{abstract}
%

\section{Introduction}
The formalism of supersymmetric quantum mechanics (SQM), introduced by Witten 
\cite{Wit81} in 1981, provides an elegant way to construct a hierarchy of
Hamiltonians with well defined relations between their spectra
\cite{Wit81,And84,Suk85a,Suk85b,Bay87}. Specifically, it allows the 
elimination or introduction of bound states \cite{Anc92} without altering 
the remaining part of the spectrum. Transformations  of SQM, which we 
will refer to  as  supersymmetric (SUSY) transformations,  are specific 
Darboux transformations \cite{Dei79,Lub86} of the Schr\"odinger equation 
and are related to the factorization method \cite{Inf51}. 
Consequently, there is a close connection to inverse scattering 
theory \cite{Suk85b,Kwo86}.\par

The SUSY-transformations of the radial Schr\"odinger equation have been 
studied in detail \cite{Bay87,Anc92}. In particular emphasis was given to 
mathematical aspects as well as applications to specific physics 
phenomena. An important application is the relationship of equivalent 
effective interactions between composite particle systems. In such systems,
because of the necessary suppression of the internal degrees of freedom, 
the resulting interactions are ambiguous \cite{Lee80,Fri81}. There is 
always a set of effective interactions leading to the same scattering 
phase shifts which sustain, however, a different number of additional
unphysical bound states as a consequence of simulating differently the
Pauli principle. \par

Baye \cite{Bay87} pointed out that 
SQM  is an elegant way to construct such phase equivalent potentials. 
Since then, more general transformations of the radial Schr\"odinger 
equation for uncoupled channels have been worked out \cite{Bay87,Anc92}
which, apart from a simple removal and an addition of a ground state, 
allow arbitrary modifications of the bound state spectrum \cite{Bay94}.\par

The concept of SUSY--transformations has been extended to 
coupled--channel systems by Amado {\em et al.} \cite{Ama88} but their 
transformations do not allow the construction of phase-equivalent 
potentials after the removal of a bound state \cite{Ama90}. 
Recently, Sparenberg and Baye \cite{Spa97} addressed this 
problem and presented  SUSY--transformations  for coupled--channel 
systems which in addition to the removal of bound states  allow the 
construction of phase-equivalent potentials. However, the status of 
SUSY-transformations for coupled-channel systems is still incomplete 
because the process of introducing a bound state has not been considered 
so far. In this article we present a  SUSY--transformation 
for the introduction of $N$ degenerate bound states in a 
system of $N$ coupled channels. \par

\section{Supersymmetric transformation}
We consider a system of $N$ coupled channels which is described by the
Schr\"odinger equation 
\begin{equation}
        H_0 \Psi_0(\epsilon ,r) \equiv 
        \left\{ -\frac{{\rm d}^2}{{\rm d}r^2} + U_0(r) \right\} 
        \Psi_0(\epsilon , r) =
        \epsilon \Psi_0 (\epsilon ,r) \, ,
\label{eq:sch}
\end{equation}
where $\epsilon = k^2 = 2mE/\hbar^2$  and $U_0(r) = 2mV_0(r)/\hbar^2$.
The potential $V_0(r)$ is an $N\times N$ matrix that 
may  include the centrifugal barrier and thresholds which may be different 
in each channel. For simplicity, we assume that the mass $m$ is 
equal for all channels (for unequal  masses, see Ref. \cite{Ama90}). 
The wave  function $\Psi_0(\epsilon ,r)=\left( \psi^1_0(\epsilon,r), \cdots ,
\psi^N_0(\epsilon ,r)\right)$ is an $N\times N$ matrix, where 
each column vector $\psi^i_0(\epsilon ,r)$ is a solution of 
Eq. (\ref{eq:sch}).\par

The SUSY transformations are based on the factorization of the 
Hamiltonian
\begin{equation}
        H_0 = A^+ A^- + \bar \epsilon \, ,
\label{eq:fac}
\end{equation}
where the energy $\bar \epsilon $ is smaller than or equal to
the ground state $\epsilon_0$ of  $H_0$ and  
\begin{equation}
        A^\pm = \pm \frac{{\rm d}}{{\rm d}r} + W(r) \, .
\label{eq:susyA}
\end{equation}
The superpotential $W(r)$ is an $N\times N$ matrix satisfying
the differential equation
\begin{equation}
        \frac{{\rm d}W}{{\rm d}r} +W^2 = U_0 - \bar \epsilon \, .
\label{eq:diffw}
\end{equation}
The supersymmetric partner Hamiltonian $H_1$ is given by
\begin{equation}
        H_1 = A^- A^+ + \bar \epsilon \quad \mbox{with} \quad
        U_1(r) = U_0(r) - 2 \frac{{\rm d}}{{\rm d}r} W(r) \, .
\label{eq:H1}
\end{equation}
The solutions of the corresponding Schr\"odinger equation,
at any energy $\epsilon $, are directly given in terms of 
$\Psi_0(\epsilon ,r)$ by
$\Psi_1(\epsilon ,r) = A^- \Psi_0(\epsilon ,r)$.
The above equations are formally equivalent to those for uncoupled 
channels for which the SUSY transformations are known in closed form 
\cite{Suk85a,Suk85b,Bay87}. In the present work we consider a SUSY 
transformation for coupled--channel systems generated via the ansatz
\begin{eqnarray}
        W(r) & = & 
        \Psi '_0(\bar \epsilon ,r) \Psi^{-1}_0(\bar \epsilon ,r) +
        \left( \Psi_0^\dagger (\bar \epsilon ,r) \right)^{-1}\Lambda
\label{eq:susynew} \\
 &\times &
        \left[ 1 + \int_a^r {\rm d}t \Psi_0^{-1}(\bar \epsilon ,t) 
\left( \Psi_0^\dagger (\bar \epsilon ,t) \right) ^{-1} \Lambda 
\right]^{-1} \Psi_0^{-1}(\bar \epsilon ,r)  
\nonumber
\end{eqnarray}
where we have used the hermiticity \cite{Ama88} of the first term of
Eq. (\ref{eq:susynew}). The constant $N\times N$ matrix $\Lambda $ 
must be symmetric and should be chosen in such a way that $W(r)$ has 
no singularities for $r>0$. The first term of Eq. (\ref{eq:susynew}) 
is a straightforward extension of the simplest expression for the
superpotential in uncoupled systems. 

The superpotential (\ref{eq:susynew}) can also be put in the 
simplest form $\Psi'_0 \Psi_0^{-1}$, where $\Psi_0$ is now the most 
general solution of the Schr\"odinger equation \cite{Spa98}. 
The associated transformation depends on the choice of 
$\Psi_0(\bar \epsilon ,r)$, the integration bound $a$, and the constant 
matrix $\Lambda $. In the present work we are concerned with choices 
that lead to the addition of a  bound state.\par

For an energy $\bar \epsilon \leq \epsilon_0$, it turns out that 
the regular solution of $H_0$ together with the bound $a=\infty $ 
is the appropriate choice. Then the transformed wave function 
$\Psi_1(\epsilon ,r)$ at the energy $\epsilon = \bar \epsilon $, is 
simply given by
\begin{eqnarray}
  \Psi_1(\bar \epsilon ,r) & = & A_0^- \Psi_0(\bar \epsilon ,r)
      \nonumber \\ & = &
    \left( \Psi_0^\dagger (\bar \epsilon ,r)\right)^{-1}\Lambda 
\nonumber \\
  & \times & \left[ 1 - \int_r^\infty {\rm d}t 
        \Psi_0^{-1}(\bar \epsilon ,t)
        \left( \Psi_0^\dagger (\bar \epsilon ,t)\right)^{-1} 
        \Lambda \right]^{-1}\, .
\label{eq:bound}
\end{eqnarray}
The regular solutions $\Psi_0$, at energies below the lowest ground state 
energy of $H_0$,  exhibit a diverging behavior for $r\to \infty $. 
Hence,  all columns of $\Psi_1(\bar \epsilon ,r)$  are exponentially 
vanishing solutions of $H_1$ at asymptotic distances. Furthermore 
$\Psi_1(\bar \epsilon ,r)$  is bounded (for  $0 < r < \infty$)  if
\begin{equation}
        \det \left( 1  - \int_r^\infty {\rm d}t 
        \Psi_0^{-1}(\bar \epsilon ,t)
        \left( \Psi_0^\dagger (\bar \epsilon ,t)\right)^{-1} \Lambda 
        \right) \not = 0 
\label{eq:det}
\end{equation}
which is the case when the matrix $\Lambda $ is chosen negative semidefinite.
This guarantees also that $U_1(r)$ is bounded except for $r=0$.\par

The behavior of $\Psi_1(\bar \epsilon ,r)$ at $r\sim 0$ requires  
more attention as it depends on the coupled--channel system considered.
Specifically, if the singularity of the potential at the origin is of
the form $\nu (\nu+1)/r^2$ with the same $\nu $ for all channels, one 
obtains the behavior 
$\lim_{r\to 0} \Psi_1(\bar \epsilon ,r) r^{-\nu }= {\rm const} \neq 0$.
Hence, for $\nu \geq 1$ the function $\Psi_1(\bar \epsilon , r)$ vanishes 
at the origin and with an appropriate choice of $\Lambda $, satisfying 
(\ref{eq:det}), each column vector of $\Psi_1(\bar \epsilon ,r)$ 
corresponds to  a bound state of $H_1$. Thus the transformation 
associated with Eq. (\ref{eq:susynew}) enables us to add $N$ degenerate 
bound states to the spectrum of $H_0$ at the energy $\bar \epsilon $. 
More precisely, analytical examples show that the degeneracy at 
$\epsilon=\bar \epsilon $ seems to depend on
the rank of the matrix $\Lambda $ \cite{Spa98}. For $\nu =0$ the function 
$\Psi_1(\bar \epsilon ,r)$ does not correspond to bound states.
The situation becomes even more intriguing, when the potential exhibits
different singularities in the coupled channels, i.e. 
$\nu_i$, $i=1,\dots ,N$. It can be shown that if the coupling vanishes 
near the origin, the $i$th component of each column vector of  
$\Psi_1(\bar \epsilon ,r)$ behaves as $r^{\nu_i}$ there and leads, with 
an appropriate choice of $\Lambda $, to the addition of up to $N$ 
degenerate bound states to the spectrum of $H_0$. \par

For a nonvanishing coupling near the origin and different $\nu_i$ the 
behavior of $\Psi_1(\bar \epsilon ,r)$ near the origin cannot be given 
in closed form. Therefore, definite conclusions about the number of 
bound states that can be added are difficult to draw. However, the 
numerical examples discussed below demonstrate that the transformation 
(\ref{eq:bound}) can also lead, in the general case, to the addition of 
up to $N$ degenerate bound states.\par
 
\section{Phase-equivalence}
The SUSY transformation generated by $W(r)$ of Eq. (\ref{eq:susynew}) 
modifies, similarly to the single channel case, the S-matrix \cite{Ama88}. 
To compensate this, another transformation is required which is 
associated with a superpotential that has the opposite sign at asymptotic 
distances \cite{Spa97}. Therefore we first perform a transformation 
generated by the 
simple superpotential (first term of Eq. (\ref{eq:susynew})) using 
asymptotically vanishing solutions $\eta_0(\bar \epsilon ,r)$ of the
Schr\"odinger equation with $H_0$ [$\lim_{r\to \infty}$ 
$\eta_0(\bar \epsilon ,r)\exp (\sqrt{\bar \epsilon }\ r)= A$, where $A$
is an $N\times N$ matrix].  
The existence of the transformed potential requires a non-vanishing 
determinant of $\eta_0(\bar \epsilon ,r)$ for all finite $r$-values. This 
can always be satisfied because $\bar \epsilon $ is assumed to be smaller 
than the deepest bound state of $H_0$. The first SUSY transformation 
modifies the S-matrix but does not change the number of bound states. In 
a second step we transform $H_1$ via the transformation mediated by the 
superpotential of Eq. (\ref{eq:susynew}).

A direct verification shows that $(\eta_0^{-1}(\bar \epsilon ,r))^\dagger$ 
is a regular solution of $H_1$. We use this solution to construct the
superpotential of the second transformation, following the principle of
the preceding section. The potential $U_2$ resulting from the two 
successive transformations, as well as the corresponding solution $\Psi_2$,
can be expressed in terms of the solutions of the initial equation only. 
They read
\begin{equation}
U_2(r)  =  U_0(r) 
- 2\frac{d}{dr}\left\{ 
\chi_2(\bar \epsilon ,r) 
\eta_0^\dagger(\bar \epsilon , r) \right\}  
\label{U2}
\end{equation}
and
\begin{equation}
\Psi_2(\epsilon ,r)  =  -(\epsilon - \bar \epsilon )\Psi_0(\epsilon ,r)
-\chi_2(\bar \epsilon ,r) 
W[\eta_0(\bar \epsilon ,r),\Psi_0(\epsilon ,r)] \, ,
\label{psi2}
\end{equation}
where 
%
\begin{equation}
\chi_2(\bar \epsilon ,r)=\eta_0(\bar \epsilon ,r) 
\Lambda \left[ 1-\int_r^\infty dt \eta_0^\dagger (\bar \epsilon ,t)
\eta_0(\bar \epsilon, t) \Lambda \right]^{-1}
\label{chi}
\end{equation}
corresponds to the added bound state(s) of $H_2$ at 
$\epsilon = \bar \epsilon$. The quantity $\Lambda $ is a constant 
$N\times N$ matrix.
 
This procedure transforms a regular solution $\Psi_0$ into a regular 
solution $\Psi_2$ as long as the singularity at the origin is sufficiently 
strong ($\nu \geq 2$). In this case the new potential $U_2$ is phase 
equivalent to $U_0$ [Eq.\ (\ref{psi2}) shows that both $\Psi_0$ and
$\Psi_2$ have the same asymptotic behaviour, up to a normalization
factor, which confirms phase equivalence], and sustains (besides the 
bound states of $H_0$) up to $N$ additional bound states at 
$\bar \epsilon $. Similarly to the single SUSY transformation, discussed
in the previous section, the degeneracy of the introduced bound states 
at $\bar \epsilon $ can be controlled by the number of non-vanishing 
eigenvalues of the constant $\Lambda $ \cite{Spa98}. 

\section{Examples} 
As a demonstration we consider an $\ell =2$ two-channel system without 
threshold and a potential of  Gau\ss {} form, 
$V_{ij} =V_{ij}^{(0)} \exp (-r^2/R^2)+\delta_{ij} 6\hbar^2/2mr^2$, 
with $R=2$\,fm, $2mc^2=938.9185$\,MeV and the depths $V_{11}^{(0)}=-100$\,MeV, 
$V_{22}^{(0)}=-60$\,MeV, and $V_{12}^{(0)}=V_{21}^{(0)}=-56.56$\,MeV. 
This potential does not sustain a bound state. A pair of degenerate 
bound states at $E=-200$\,MeV has been introduced using the twice iterated
SUSY transformation. The new potentials are shown in Fig.\,1 together with 
the corresponding wave functions of the two added bound states. It is seen 
that there is a clear spatial separation between the two degenerate states 
which is also reflected in the potential matrix. Choosing $\Lambda $ 
of rank one results in the introduction of only one bound state at 
$\bar \epsilon $.
This leads to the potential and bound state wave functions displayed in
Fig.\,2. It must be emphasized that the potentials of Fig.\,1 and Fig.\,2 are
phase equivalent.\par

The introduction of bound states via the twice iterated SUSY transformation 
(\ref{U2}-\ref{chi}) is also possible for systems where channels with 
different quantum numbers of orbital angular momentum are coupled. Of 
particular interest is the coupling of an s-channel ($\ell_1=0$) and a
d-channel ($\ell_2=2$) when the associated potential has only in the 
d-channel an $1/r^2$-singularity at the origin. In such a system the strength
of the singularity suffices for the introduction of one additional bound
state via the SUSY transformation (\ref{U2}-\ref{chi}). The transformed
potential has no $r^{-2}$ singularity at the origin which, taking into 
account the centrifugal term of the d-channel explicitly, implies an 
attractive $1/r^2$-contribution to the residual potential. We have 
verified this special case numerically
assuming the previous two-channel system with slight modifications, i.e.
$\ell_1=0, \ell_2=2$ and 
$V_{ij}=V_{ij}^{(0)} \exp (-r^2/R^2)+\delta_{2j}6\hbar^2/2mr^2$. 
This system has a bound state at $E=-29.45$\ MeV. 
The introduction of an additional bound state at $E=200$\ MeV leads to the
potential and bound state wave function displayed in Fig.\,3. The transformed
potential confirms the previous discussion and its phase equivalence has been
verified numerically. On the contrary, for a $\Lambda $ matrix of rank 2, 
we have verified that the final potential is not phase equivalent 
to $U_0$.\par

As a realistic example we consider the $^3$S$_1$-$^3$D$_1$ partial waves
of the nucleon--nucleon (NN) system which are coupled by a tensor 
potential that sustains a single physical bound state 
at $E_d=-2.22$\,MeV (the deuteron). Within the quark model the NN 
force is an effective interaction of a composite particle system and 
may give rise to  Pauli--forbidden states  which are simulated by deep 
{\it unphysical} bound states in local potential models \cite{Lee80}.
Neudatchin {\em et al.} \cite{Neu73} were the first to use the concept 
of forbidden states for the NN potential. A more detailed study led to 
the well known Moscow potential which reproduces the scattering data up 
to $400$\,MeV \cite{Kuk98}. Elimination of the additional unphysical 
bound state via the twice iterated SUSY transformation of Sparenberg 
and Baye \cite{Spa97} leads to a phase equivalent potential having a 
close  similarity to the standard NN interactions derived in the 
meson-exchange picture. Qualitatively the same result has been obtained 
by Leeb {\em et al.} \cite{Lee94} by applying a numerical SUSY 
transformation to an early version of the Moscow potential.\par

In the present work, we apply the procedure described by Eqs. 
(\ref{U2}-\ref{chi}) in 
the opposite direction. Starting from the Reid soft core potential (RSC) 
\cite{Rei68} for the $^3$S$_1$-$^3$D$_1$ channel, we would like to generate 
a series of NN potentials of Moscow-type where an additional bound state 
has been introduced at different energies $\bar \epsilon $. 
In Fig.\,4 we show a series of phase equivalent NN--potentials obtained by 
this procedure. Variation of the matrix $\Lambda$ results in 
modifications of the $r$-dependence while the energy of the added bound
state is reflected in the depth of the potentials. The repulsive core
of the effective potential $V_{22}(r)$ is a consequence of the repulsive
core of the RSC potential. It must be emphasized that the obtained 
potentials are not exactly of the same nature as the Moscow potential 
since the central part of their D-channel component has an $r^{-2}$ 
attractive core. 

\section{Conclusions}
In summary, in this work we have  presented a closed form expression
for a SUSY transformation which enables us to introduce single or degenerate
bound states in a coupled channel system. The novel transformation works
for a general hermitean $N$-channel system with and without thresholds.
The method has been successfully tested for the case of two coupled partial 
waves in the NN system where a whole set of phase equivalent NN potentials 
of Moscow-type has been generated. The method has also been successfully 
applied to two--channel systems with potentials of Gau\ss  {} form.

\acknowledgements
The work has been supported by {\it Fonds zur F\"orderung der
wissenschaftlichen Forschung, \"Osterreich}, project number P10467-PHY
and the Foundation for Research Development of South Africa. 
J.-M. S. is supported by the National Fund for Scientific Research,
Belgium. This text presents research results of the Belgian program
P4/18 on interuniversity attraction poles initiated by the Belgian-state
Federal Services for Scientific, Technical and Cultural Affairs.


\vfill
\newpage

\section*{Figure Captions}

\noindent {\bf Figure 1}\\
Introduction of two degenerate bound states at $E=-200$\,MeV into a
two-channel system with Gau\ss ian potential (see text) at $\ell =2$ 
using $\Lambda_{11}=-5\cdot 10^4$, $\Lambda_{22}=-5\cdot 10^6$, 
$\Lambda_{12}=\Lambda_{21}=0$ and $\eta_0(\bar \epsilon ,r)$
associated with $A_{11}=A_{12}=1$ and $A_{21}=-A_{22}=0.5$. The upper 
graph shows the matrix elements of the potential; $V_{11}(r)$ (solid line), 
$V_{22}(r)$ (dashed line) and $V_{12}(r)=V_{21}(r)$ (dotted line). 
The two lower graphs show the components of the wave functions (solid line 
for the first channel, dashed line for the second channel) of the two added
bound states.

\vspace{0.5cm}
\noindent {\bf Figure 2}\\
The same system as shown in Fig.\,1 but entering only one bound state at
$E=-200$\,MeV using $\Lambda_{11}=-5\cdot 10^4$, 
$\Lambda_{12}=\Lambda_{21}=\Lambda_{22}=0$ and the same
$\eta_0$. For the notation see Fig.\,1

\vspace{0.5cm}
\noindent {\bf Figure 3}\\
Introduction of a bound state at $E=-200$\,MeV into a two-channel system 
of coupled s- and d-partial waves with Gau\ss ian potential (see text) 
using $\Lambda_{11}=-5\cdot 10^4$, 
$\Lambda_{12}=\Lambda_{21}=\Lambda_{22}=0$ and 
$\eta_0(\bar \epsilon ,r)$ 
associated with $A_{11}=A_{12}=1$, $A_{21}=-A_{22}=0.5$. The upper graph 
shows the matrix elements of the potential; $V_{11}(r)$ (solid line), 
$V_{22}(r)$ (dashed line) and $V_{12}(r)=V_{21}(r)$ (dotted line). 
The graph at bottom shows the components of the wave function (solid line 
for the first channel, dashed line for the second channel) of the added 
bound state.

\vspace{0.5cm}
\noindent {\bf Figure 4}\\
NN potentials for the $^3$S$_1$-$^3$D$_1$-channel, phase equivalent to the RSC 
generated by the twice iterated SUSY transformation  (\ref{eq:susynew}). 
The unphysical bound states are at $E=-100$\,MeV (long dashed line), 
$E=-200$\,MeV (dashed line), $E=-400$\,MeV (dotted line) and 
$E=-800$\,MeV (solid line). The constant matrix $\Lambda $ was fixed to 
$\Lambda_{22}=-5\cdot 10^{6} $, 
$\Lambda_{11}=\Lambda_{12}=\Lambda_{21}=0 $ and 
$\eta_0(\bar \epsilon ,r)$ associated with $A_{11}=A_{12}=1$ and
$A_{21}=-A_{22}=0.5$.



\begin{center}

\begin{figure}[htb]
\centerline{\epsfig{file=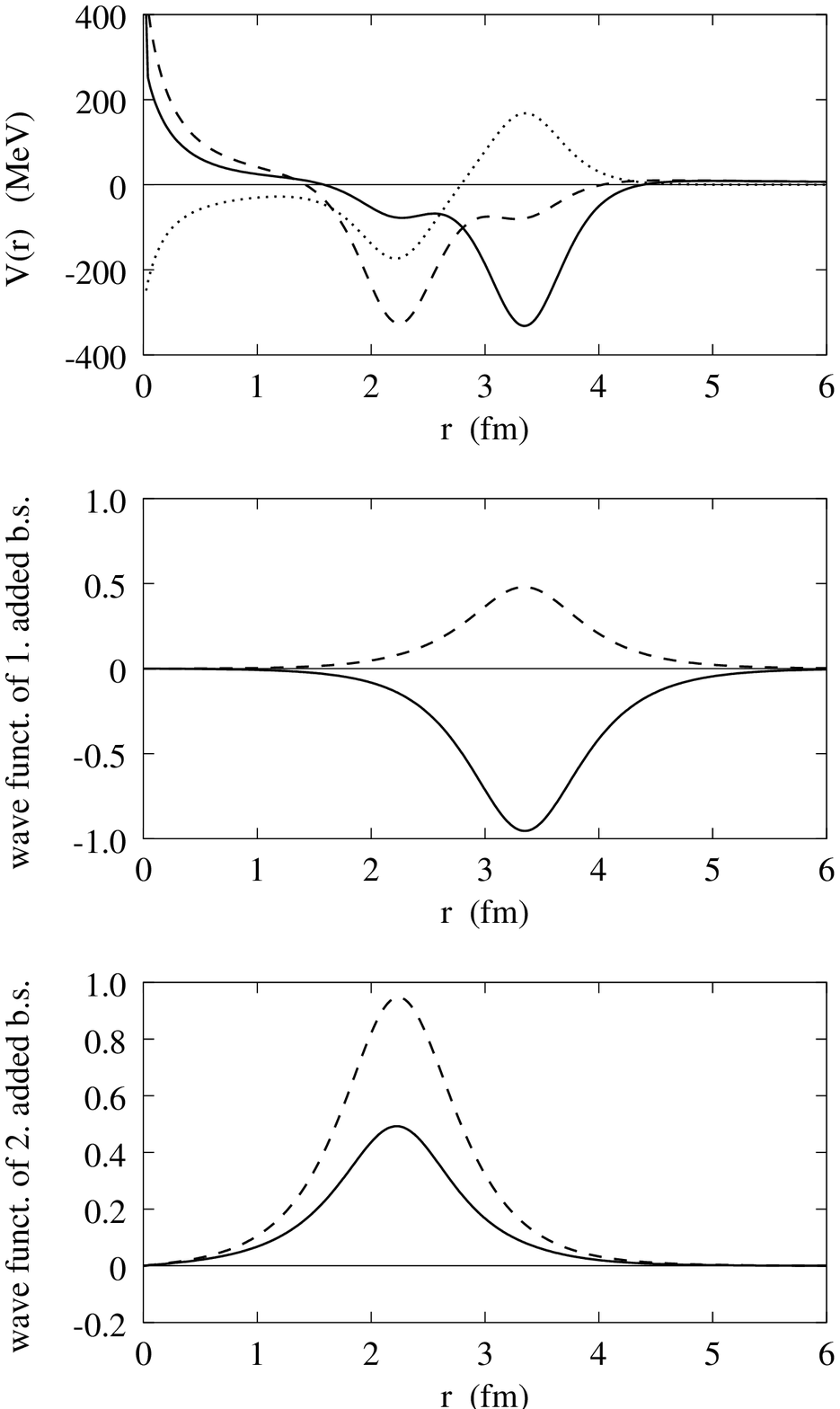,width=12.0cm}}
\end{figure}

\vspace{0.5cm}

{\large {\bf Figure 1}}

\vfill
\newpage

\begin{figure}[htb]
\centerline{\epsfig{file=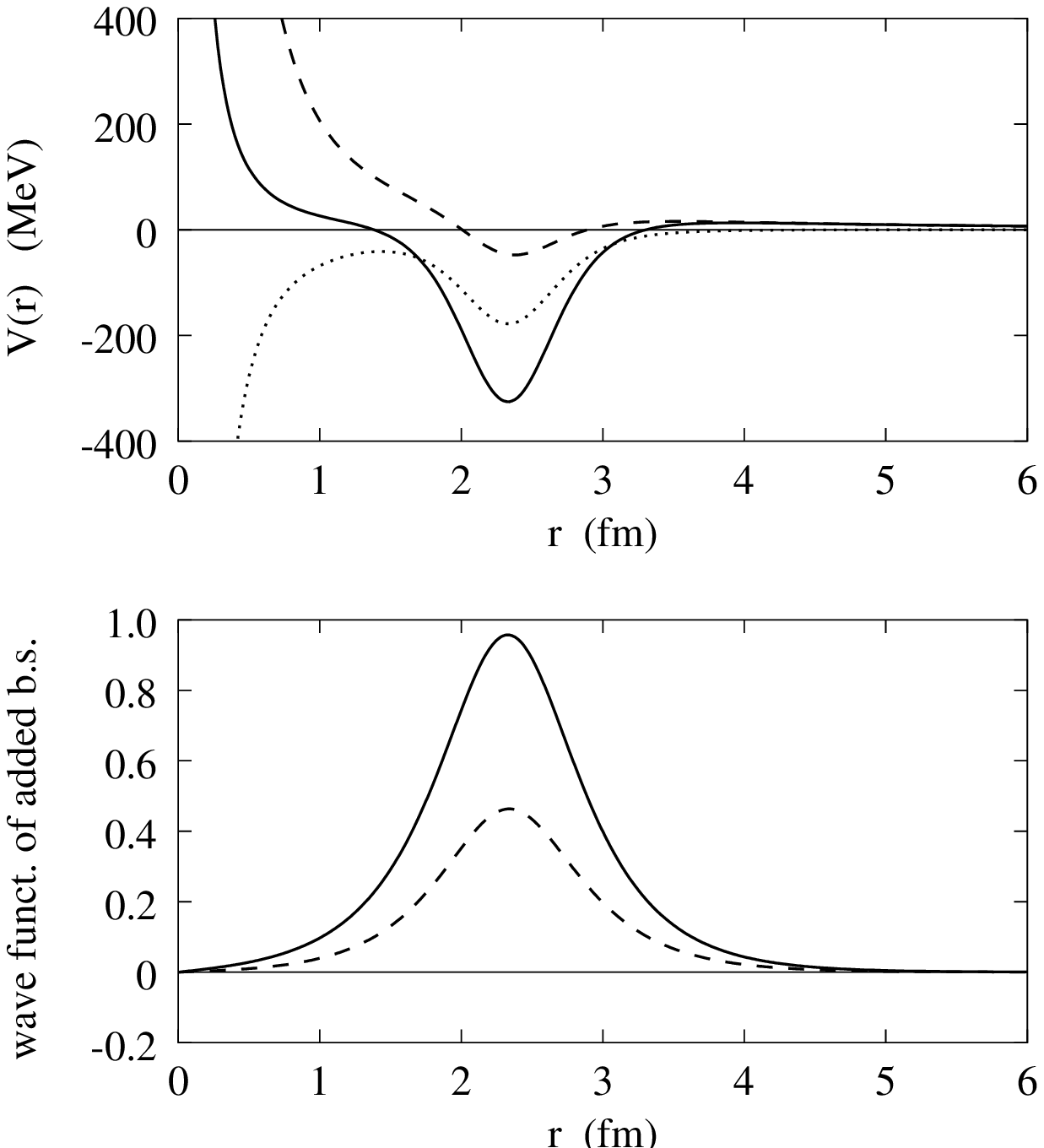,width=12cm}}
\end{figure}

\vspace{1cm}

{\large {\bf Figure 2}}

\vfill
\newpage

\begin{figure}[htb]
\centerline{\epsfig{file=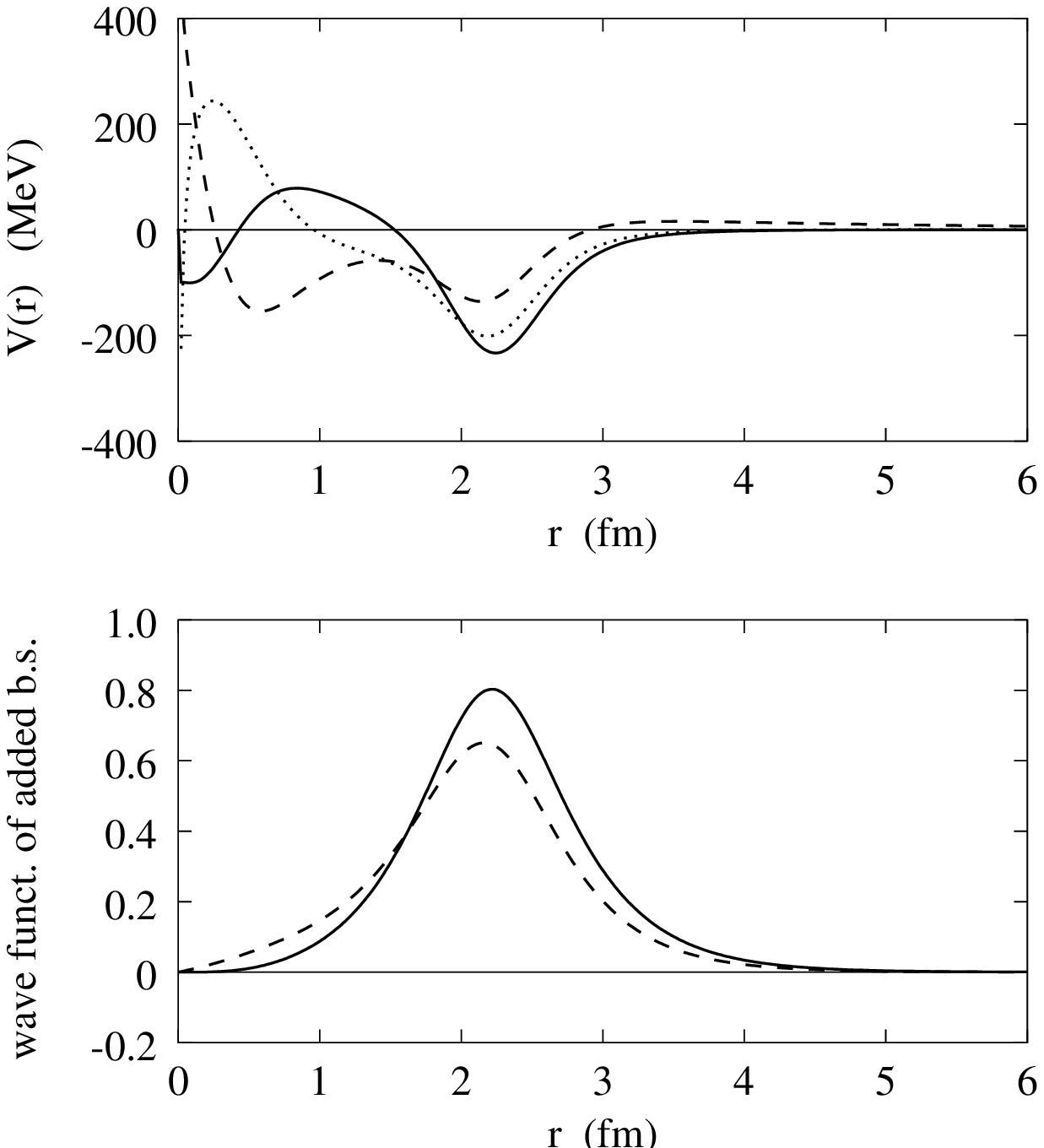,width=12.0cm}}
\end{figure}

\vspace{1cm}

{\large {\bf Figure 3}}

\vfill
\newpage

\begin{figure}[htb]
\centerline{\epsfig{file=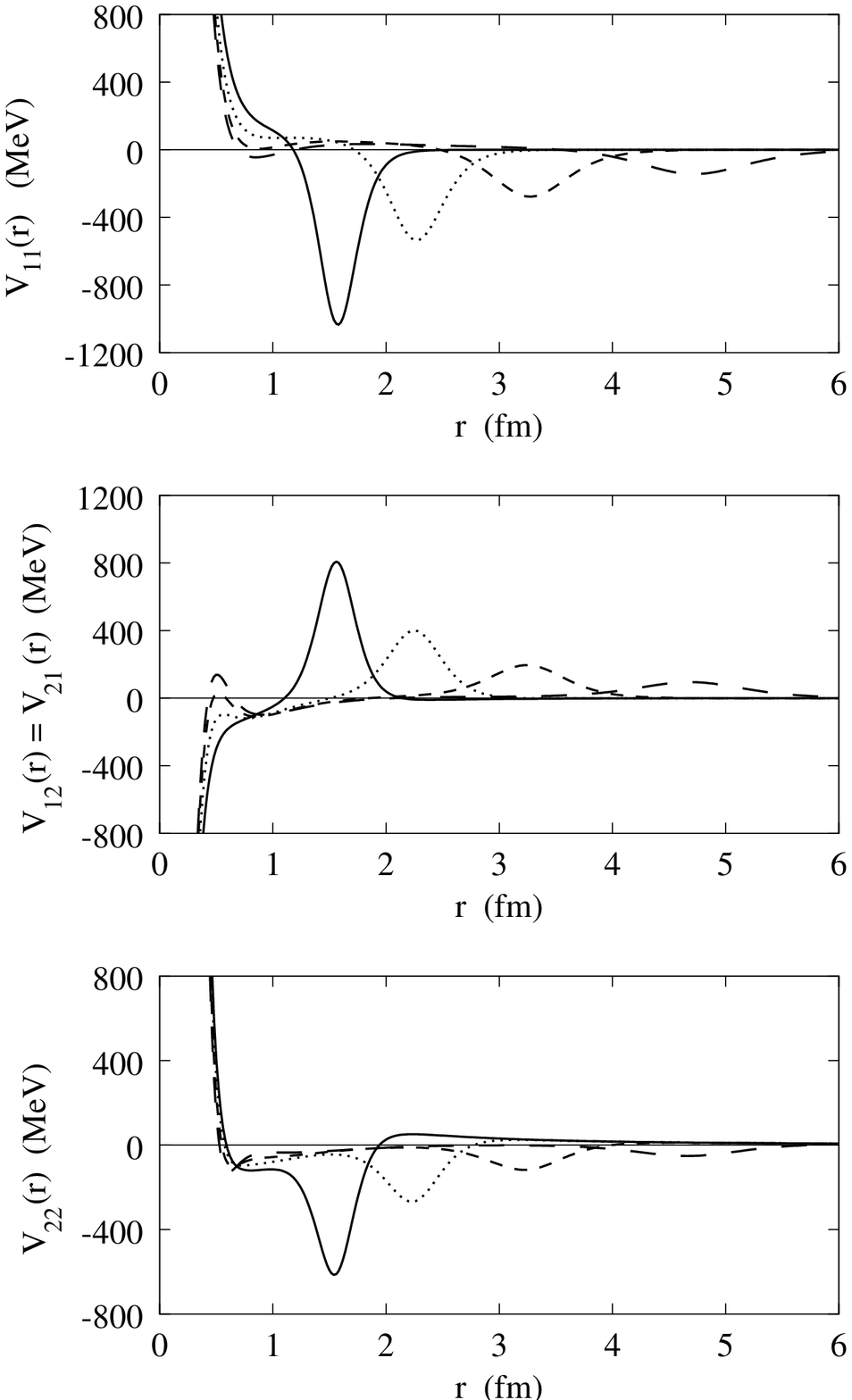,width=12.0cm}}
\end{figure}

\vspace{1cm}

{\large {\bf Figure 4}}

\end{center}

%

\end{document}